\documentclass[a4paper]{article}
\usepackage[affil-it]{authblk}
\usepackage[left=20mm,right=20mm,top=20mm,bottom=23mm]{geometry}
\usepackage{setspace}
\usepackage{amsmath}
\usepackage{subfig}
\usepackage{graphicx}
\usepackage{multirow}

\newcommand{\Br}{\mathrm{Br}}
\newcommand{\e}{\mathrm{e}}

\begin{document}

\date{}

\title{New Physics at 1 TeV?}

\author[1,2]{S. I. Godunov    \thanks{sgodunov@itep.ru}}
\author[3]{A. N. Rozanov}
\author[1,4,5]{M. I. Vysotsky \thanks{vysotsky@itep.ru}}
\author[1,5]{E. V. Zhemchugov \thanks{zhemchugov@itep.ru}}
\affil[1]{\small Institute for Theoretical and Experimental Physics,
117218, Moscow, Russia}
\affil[2]{\small Novosibirsk State University,
630090, Novosibirsk, Russia}
\affil[3]{\small The Center for Particle Physics of Marseilles,
CPPM-IN2P3-CNRS-AMU, F-13288, Marseille, France}
\affil[4]{\small Moscow Institute of Physics and Technology, 141700,
Dolgoprudny, Moscow Region, Russia}
\affil[5]{\small Moscow Engineering Physics Institute,
115409, Moscow, Russia}

\maketitle

\begin{abstract}
 If decays of a heavy particle $S$ are responsible for the diphoton excess with
 invariant mass 750~GeV observed at the 13~TeV LHC run, it can be easily
 accomodated in the Standard Model. Two scenarios are considered: production in
 gluon fusion through a loop of heavy isosinglet quark(s) and production in
 photon fusion through a loop of heavy isosinglet leptons. In the second case
 many heavy leptons are needed or\slash{}and they should have large electric
 charges in order to reproduce experimental data on $\sigma_{pp \to SX} \cdot
 \Br(S \to \gamma \gamma)$.
\end{abstract}

\section{Introduction}

ATLAS and CMS collaborations recently announced a small enhancement over smooth
background of two photon events with invariant mass
750~GeV~\cite{atlas-750gg,cms-750gg}. Though statistical significance of this
enhancement is not large (within 3 standard deviations), it induced a whole
bunch of theoretical papers devoted to its interpretation. The reason
for this explosive activity is clear: maybe the Standard Model of Particle
Physics is changed at one TeV scale, and we are witnessing the first sign of
this change.

Let us suppose that the observed enhancement is due to the $\gamma \gamma$ decay
of a new particle. Then it should be a boson with spin different from one; the
simplest possibility is a scalar particle $S$ with $m_S = 750$~GeV. Since it
decays to two photons, it should be an $SU(3)_\text{c}$ singlet, and in
$pp$-collisions at the LHC it can be produced in gluon-gluon fusion through the
loop of colored particles and in photon-photon fusion through the loop of
charged particles. Let us suppose that particles propagating in the loops are
heavy, and $S$ decays to them are kinematically forbidden.\footnote{
 In the opposite case $\Br(S \to \gamma \gamma)$ reduces significantly which
 makes $S \to \gamma \gamma$ decays unobservable at the LHC.
}
Production cross section is evidently larger in the case of gluon fusion,
however $S \to \gamma \gamma$ branching ratio is suppressed in this case since
$S \to g g$ decay dominates.

We suppose that the particles propagating in the loop are Dirac fermions, so
they have tree level masses, and that they are $SU(2)_\text{L}$ singlets.
Nonzero hypercharges provide couplings of these particles with photon and
$Z$-boson. These particles can be quark(s) (color triplets) $T_i$ or lepton(s)
(color singlets) $L_i$. They couple with $S$ by Yukawa interactions with
coupling constants $\lambda^i_T$ and $\lambda^i_L$ correspondingly.

In Section~\ref{s:quarkophilic} we will consider $S$ production and decay in the
model with extra heavy quark(s), in which gluon fusion dominates $S$ production;
in Section~\ref{s:leptophilic} we will consider the model with extra heavy
lepton(s), where $S$ production occur in photon fusion, and $S \to \gamma
\gamma$ decay dominates.

\section{Quarkophilic $S$}

\label{s:quarkophilic}

In the case of one heavy quark $T$ the following terms should be added to the
Standard Model lagrangian:
\begin{equation}
 \Delta \mathcal{L}
 = \tfrac12 (\partial_\mu S)^2
 - \tfrac12 m_S^2 S^2
 + \bar T \gamma_\mu (
      \partial_\mu
    - \tfrac{i}{2} g_s A_\mu^i \lambda_i
    - i g' \tfrac{Y_T}{2} B_\mu
   ) T
 + m_T \bar T T
 + \lambda_T \bar T T S,
 \label{lagrangian-t}
\end{equation}
where $A_\mu^i$ and $B_\mu$ are gluon and $U(1)$ gauge fields respectively, and
$\lambda_i$ are Gell-Mann matrices. $S$ coupling with gluons is generated by
the $T$-quark loop:
\begin{equation}
 M_{gg}
 = \frac{\alpha_s}{6 \pi} \frac{\lambda_T}{m_T} F(\beta)
   G_{\mu \nu}^{(1)} G_{\mu \nu}^{(2)} S,
 \label{S->gg-amplitude}
\end{equation}
where $\beta = (2 m_T / m_S)^2$,
\begin{equation}
 F(\beta)
 = \frac32 \beta
   \left[ 1 - (\beta - 1) \arctan^2 \frac{1}{\sqrt{\beta - 1}} \right],
\end{equation}
and $F(\beta) \to 1$ for $m_T \gg m_S$.

\begin{figure}
 \centering
 \includegraphics{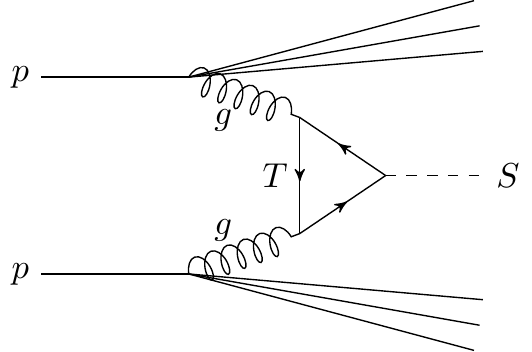}
 \caption{Feynman diagram of $S$ production.}
 \label{fig:S-production}
\end{figure}

Inclusive cross section of $S$ production in $pp$ collision at the LHC through gluon
fusion is given by:
\begin{equation}
 \sigma_{pp \to SX}
 = \frac{\alpha_s^2}{576 \pi}
   \left( \frac{\lambda_T}{m_T} \right)^2
   \lvert F(\beta) \rvert^2
   m_S^2
   \left. \frac{dL_{gg}}{d \hat s} \right\rvert_{\hat s = m_S^2},
 \label{pp->SX-gluons}
\end{equation}
where the so-called gluon-gluon luminosity is given by the integral over gluon
distributions:
\begin{equation}
 \frac{dL_{gg}}{d \hat s}
 = \frac{1}{s}
   \int\limits_{\ln \sqrt{\tau_0}}^{-\ln \sqrt{\tau_0}}
    g(\sqrt{\tau_0} \e^y, Q^2)
    g(\sqrt{\tau_0} \e^{-y}, Q^2)
   d y,
 \label{gg-luminosity}
\end{equation}
$\tau_0 = \hat s / s$, $s = (13 \text{ TeV})^2$, and we use $Q^2 = m_S^2$. In
Fig.~\ref{fig:S-production} the corresponding Feynman diagram is shown.
Integrating gluon distributions from~\cite{mmht} for $\sqrt{\hat s} = 750\text{
GeV}$, $\sqrt{s} = 13\text{ TeV}$, we get $dL_{gg} / d \hat s \approx 4.0$~nb,
$m_S^2 \; dL_{gg} / d \hat s \approx (1 / 0.69\text{ nb}) \cdot 4.0 \text{ nb}
\approx 5.8$.  At $\sqrt{s} = 8$~TeV for $\sqrt{\hat s} = 750$~GeV the
luminosity $dL_{gg} / d \hat s$, and therefore cross
section~\eqref{pp->SX-gluons}, is $4.6$ times smaller. In order to take into
account gluon loop corrections,~\eqref{pp->SX-gluons} should be multiplied by
the so-called $K$-factor which is close to 2 for $\sqrt{s} = 13$~TeV, according
to~\cite{djouadi} (see also Fig. 2 in~\cite{harlander}).

In this way for $m_T = m_S$ and $\lambda_T = 1$, substituting $\alpha_s(m_S) =
0.090$, we obtain:
\begin{equation} 
 \sigma_{pp \to SX} \approx 41\text{ fb},
 \label{pp->SX-xsection-value}
\end{equation}
which should be multiplied by $\Br(S \to \gamma \gamma)$ in order to be compared with
experimental observations~\cite{atlas-750gg,cms-750gg}. Total width of $S$ is
dominated by the $S \to gg$ decay, and from~\eqref{S->gg-amplitude} we get:
\begin{equation}
 \Gamma_{S \to gg}
 = \left( \frac{\alpha_s}{6 \pi} \right)^2
 \cdot 8 \frac{m_S^3 \lambda_T^2}{16 \pi m_T^2} \lvert F(\beta) \rvert^2
 \approx 3.1\text{ MeV},
\end{equation}
four orders of magnitude smaller than the 45~GeV width which (maybe) follows
from the preliminary ATLAS data. Thus we conclude that for the models we
consider, $S$ width should be much smaller than 45~GeV. Let us note that CMS
data prefer narrow $S$; see also~\cite{buckley}.

\begin{figure}[b]
 \centering
 \includegraphics{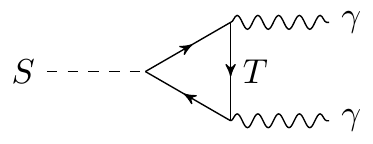}
 \caption{Feynman diagram of $S \to \gamma \gamma$ decay.}
 \label{fig:S-decay}
\end{figure} 
$T$-quark loop contributes to $S \to \gamma \gamma$ decay as well (see
Fig.~\ref{fig:S-decay}). The corresponding matrix element equals
\begin{equation}
 M_{\gamma \gamma}
 = \frac{\alpha}{3 \pi} \frac{\lambda_T}{m_T} F(\beta)
   F_{\mu \nu}^{(1)} F_{\mu \nu}^{(2)}
 \cdot 3_\text{c} Q_T^2,
\end{equation}
where the factor $3_c$ corresponds to the three colors, and $Q_T$ is the
$T$-quark electric charge. For $\gamma \gamma$ width we get:
\begin{equation}
 \Gamma_{S \to \gamma \gamma}
 = \left( \frac{\alpha}{3 \pi} \right)^2
   (3_\text{c} Q_T^2)^2
   \frac{m_S^3 \lambda_T^2}{16 \pi m_T^2} \lvert F(\beta) \rvert^2
   \approx 22\text{ keV},
 \label{S->2gamma-width}
\end{equation}
and
\begin{equation}
 \Br(S \to \gamma \gamma)
 \approx \left( \frac{\alpha}{\alpha_s} \right)^2
         \frac{(3_\text{c} Q_T^2)^2}{2}
 \approx 0.0070,
 \label{S->2gamma-branching}
\end{equation}
where we substituted $Q_T = 2/3$ and $\alpha = 1/125$.\footnote{
 Fine structure constant should be substituted by its running value at $q^2 =
 m_S^2$, $\alpha(m_S^2) = 1/125$.
}
Finally, from~\eqref{S->2gamma-branching} and~\eqref{pp->SX-xsection-value} we
obtain:
\begin{equation}
 \sigma_{pp \to SX} \cdot \Br(S \to \gamma \gamma) \approx 0.28\text{ fb}.
 \label{pp->SX*branching-value}
\end{equation}
Experimental data provides a value approximately 36 times larger:
\begin{equation}
 [\sigma_{pp \to SX} \cdot \Br(S \to \gamma \gamma)]_\text{exp}
 \approx 10\text{ fb},
 \label{pp->SX-experimental-value}
\end{equation}
since with $3\text{ fb}^{-1}$ luminosity collected by each collaboration at
13~TeV and effectivity of $\gamma \gamma$ registration $\varepsilon \approx
0.5$~\cite{atlas-750gg} they see about 15 events each.

In order to reproduce experimental result~\eqref{pp->SX-experimental-value} we
should suppose that six $T$-quarks exist. In this case $\Gamma_{S \to gg} = 36
\cdot 3.1\text{ MeV} \approx 110\text{ MeV}$, $\Br(S \to \gamma \gamma)$ remains
the same, while the cross section of $S$ production~\eqref{pp->SX-xsection-value}
should be multiplied by the same factor 36,
and~\eqref{pp->SX-experimental-value} is reproduced.\footnote{
 If at one TeV scale we have a ``mirror image'' of the Standard Model with three
 vector-like generations of quarks and leptons, then experimental
 result~\eqref{pp->SX-experimental-value} will be reproduced.
}

\begin{figure}
 \centering
 \includegraphics{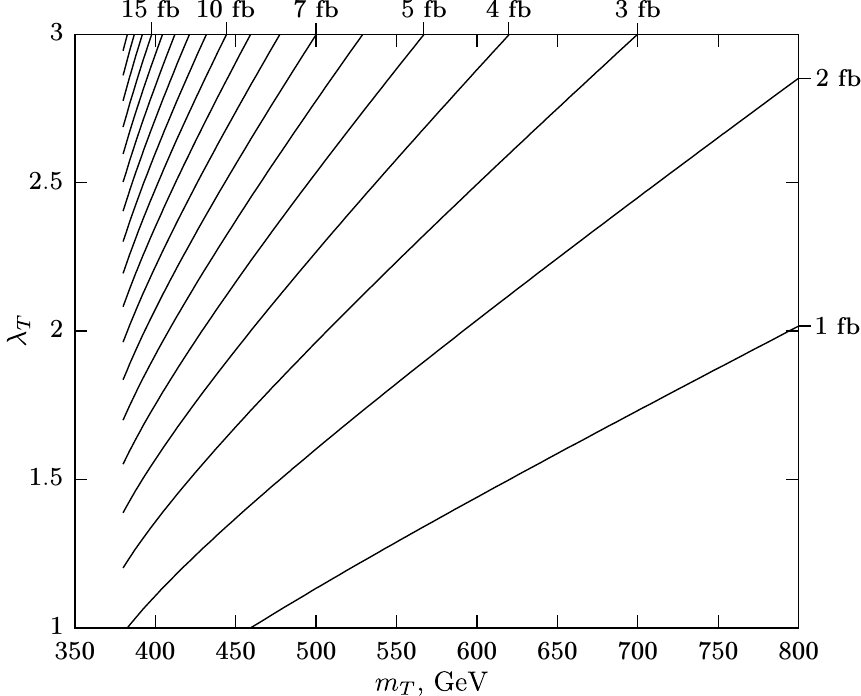}
 \caption{Contour plot of $\sigma_{pp \to SX} \cdot \Br(S \to \gamma
 \gamma)$.}
 \label{fig:pp->SX}
\end{figure}

However, unappealing multiplication of the number of $T$-quarks can be avoided. For $m_T =
400$~GeV we have $F(\beta) = 1.36$ and $\sigma_{pp \to SX} \cdot \Br(S \to
\gamma \gamma)$ is $5.7$ times larger than what is given
in~\eqref{pp->SX*branching-value}. Thus for $\lambda_T = 2.5$ we reproduce
the experimental number.\footnote{
 As far as $\lambda_T^2 / 4 \pi$ is a parameter of perturbation theory, this value
 of $\lambda_T$ is close to the maximum allowed value in order for the
 perturbation theory to make sense.
} In Figure~\ref{fig:pp->SX} isolines of the product
$\sigma_{pp \to SX} \cdot \Br(S \to \gamma \gamma)$ are shown on $(\lambda_T,
m_T)$ plot.

In the following we consider the model with one additional quark $T$ and
\begin{equation}
 m_T = 400\text{ GeV}, \  \lambda_T = 2.5.
 \label{quark-parameters}
\end{equation}

$S$ can mix with the Standard Model Higgs boson due to renormalizable
interaction term $\mu \Phi^\dagger \Phi S$, where $\Phi$ is the Higgs
isodoublet. Such an extension of the Standard Model was studied in our recent
paper~\cite{isosinglet}. Doublet admixture in the 750~GeV boson wave function
results in tree level decays $S \to WW$, $ZZ$, $t\bar t$ and $hh$, where $h$ is
the 125~GeV Higgs boson. According to Eqs.~(16)--(20) from~\cite{isosinglet},
the sum of these widths equals approximately $\sin^2 \alpha \cdot m_S^3 / 8 \pi
v_\Phi^2 \approx \sin^2 \alpha \cdot 300$~GeV, where $\alpha$ is the mixing
angle, and $v_\Phi = 246$~GeV is the Higgs boson vacuum expectation value. Ratio
of partial widths at small $\alpha$ is
\begin{equation}
 \Gamma_{S \to WW} : \Gamma_{S \to ZZ} : \Gamma_{S \to hh}
 \approx 2 : 1 : 1.
\end{equation}
As a result, $S$ width grows and $\Br(S \to \gamma \gamma)$ diminishes
correspondingly. Thus, experimental result~\eqref{pp->SX-experimental-value}
will not be reproduced. To reduce this effect we should make the mixing angle
$\alpha$ small enough. For example, for $\sin \alpha < 1/150$ we obtain at most
12~MeV (or 11\%) increase of the width of $S$, which is acceptable.  According
to Eq.~(7) from~\cite{isosinglet},
\begin{equation}
 \sin \alpha \approx \frac{\lvert\mu\rvert v_\Phi}{m_S^2},
\end{equation}
and it is less than $1/150$ for $\lvert \mu \rvert$ below 15~GeV.

Let us check if $S \to ZZ$ decays do not exceed experimental bounds on their
relative probability obtained at 13 and 8~TeV at the LHC. Since $\Br(S \to ZZ)$
is below $2.3 \cdot 10^{-2}$, we obtain
\begin{equation}
 [ \sigma_{pp \to SX} \cdot \Br(S \to ZZ) ]_{13\text{ TeV}} < 33\text{ fb},
\end{equation}
well below experimental upper bound which, according to Fig.~11
from~\cite{atlas-4l}, equals $4\text{ fb} / (\Br(Z \to 4 \ell))^2 = 400\text{ fb}$
at $2 \sigma$ (see also~\cite{franceschini}). Gluon-gluon luminosity is $4.6$
times smaller at 8~TeV, so we get
\begin{equation}
 [ \sigma_{pp \to SX} \cdot \Br(S \to ZZ)]_{8\text{ TeV}} < 9.0 \text{ fb},
\end{equation}
which should be compared with 60~fb experimental upper bound (Fig.~12
from~\cite{atlas-zz}).

More stringent upper bound comes from the search of $S \to hh$
decays~\cite{atlas-hh} and equals 40~fb, while in our case the cross section
equals 10~fb.

Since as it has just been written above, at $\sqrt{s} = 8$~TeV the gluon-gluon
luminosity is $4.6$ times smaller that at $\sqrt{s} = 13$~TeV, the CMS bound from
Run~1~\cite{cms-8TeV-bound}
\begin{equation}
 [\sigma_{pp \to SX} \Br(S \to \gamma \gamma)]_{8\text{ TeV}} < 1.5\text{ fb}
\end{equation}
is (almost) not violated in the model considered.

It is natural to suppose that $T$-quark mixes with $u$-, $c$-, and $t$-quark
which makes it unstable. To avoid LHC Run~1 bounds on $m_T$ following from the
search of the decays $T \to W b$, $T \to Z t$ and $T \to H
t$~\cite{tt-bound-1,tt-bound-2,tt-bound-3} which exclude $T$-quark with mass
below 700~GeV, we suppose that $T-t$ mixture is small, and $T$-quark mixing with
$u$- and $c$-quarks dominates. In this case bounds~\cite{tt-bound-1, tt-bound-2,
tt-bound-3} are avoided~\cite{buchkremer}.

Concerning $S$ decays, let us note that the dominant $S \to gg$ decay is
hidden by the two jets background produced by strong interactions. At 8~TeV LHC
energy the following upper bound was obtained~\cite{dijet-background}:
\begin{equation}
 [\sigma_{pp \to SX} \cdot \Br(S \to gg)]_{8\text{ TeV}}^\text{exp}
 < 30 \text{ pb}.
 \label{quark-8TeV-bound}
\end{equation}
In our model $\Br(S \to gg) \approx 1$. From Eq.~\eqref{pp->SX-gluons}, using
gluon-gluon luminosity at $\sqrt{s} = 8$~TeV, parameters from
Eq.~\eqref{quark-parameters}, and $K$-factor $2.5$~\cite{djouadi},
\cite{harlander}, we get
\begin{equation}
 [\sigma_{pp \to SX}]^\text{theor} \approx 0.39\text{ pb},
 \ 
 \Br(S \to gg) \approx 1,
\end{equation}
two orders of magnitude smaller than the upper bound~\eqref{quark-8TeV-bound}.

Three modes of $S$ decays to neutral vector bosons do exist and have the
following hierarchy:
\begin{equation}
 \Gamma_{S \to \gamma \gamma} : \Gamma_{S \to Z \gamma} : \Gamma_{S \to ZZ}
 = 1 : 2 (s_W / c_W)^2 : (s_W / c_W)^4,
 \label{widths-ratio}
\end{equation}
where $s_W$ $(c_W)$ is the sine (cosine) of electroweak mixing angle.\footnote{
In~\eqref{widths-ratio} we suppose that mixing of $S$ with Higgs doublet is
 negligible; in the opposite case $\Gamma_{S \to ZZ}$ can exceed $\Gamma_{S \to
 \gamma \gamma}$.
} Thus if $S \to \gamma \gamma$ decays will be observed in future Run~2 data, $S
\to \gamma Z$ and $S \to ZZ$ decays should be also looked for.

If the existence of $S$ will be confirmed with larger statistics at the LHC,
then it can be studied at $e^+ e^-$-colliders as well. For the cross section of
two-photon $S$ production in the reaction $e^+ e^- \to e^+ e^- S$, according
to~\cite[Eq.~(48.47)]{pdg},~\cite{ginzburg}, we have:
\begin{equation}
 \sigma_{ee \to eeS} (s)
 = \frac{8 \alpha^2}{m_S^3} \Gamma_{S \to \gamma \gamma}
   \left[
      f \left( \frac{m_S^2}{s} \right)
      \left( \ln \left( \frac{m_T^2 s}{m_e^2 m_S^2} \right) - 1 \right)^2
    - \frac13 \ln^3 \left( \frac{s}{m_S^2} \right)
   \right],
 \label{ee->eeS-xsection}
\end{equation}
where
\begin{equation}
 f(z) = \left( 1 + \tfrac12 z \right)^2 \ln \tfrac{1}{z}
      - \tfrac12 (1 - z) (3 + z),
 \label{ee->eeS-f}
\end{equation}
and $\Gamma_{S \to \gamma \gamma}$ is given in Eq.~\eqref{S->2gamma-width}. For
$e^+ e^-$ collider CLIC with $s = (3\text{ TeV})^2$, substituting in
Eqs.~\eqref{S->2gamma-width}, \eqref{ee->eeS-xsection} $\lambda_T = 2.5$, $m_T =
400$~GeV, $\alpha(m_S^2) = 1/125$, and $F(\beta) = 1.36$ we obtain:
\begin{equation}
 \sigma_{ee \to eeS}^\text{CLIC} \approx 0.46\text{ fb}.
\end{equation}
With projected CLIC luminosity $L = 6 \cdot 10^{34} / (\text{cm}^2 \cdot
\text{sec})$~\cite[p.~393]{pdg}, during one accelerator year ($t = 10^7$~sec)
about 300 $S$ resonances should be produced.

\section{Leptophilic $S$}

\label{s:leptophilic}

\begin{figure}
 \centering
 \subfloat[Luminosities for $\sqrt{s} = 13$~TeV.]{\includegraphics{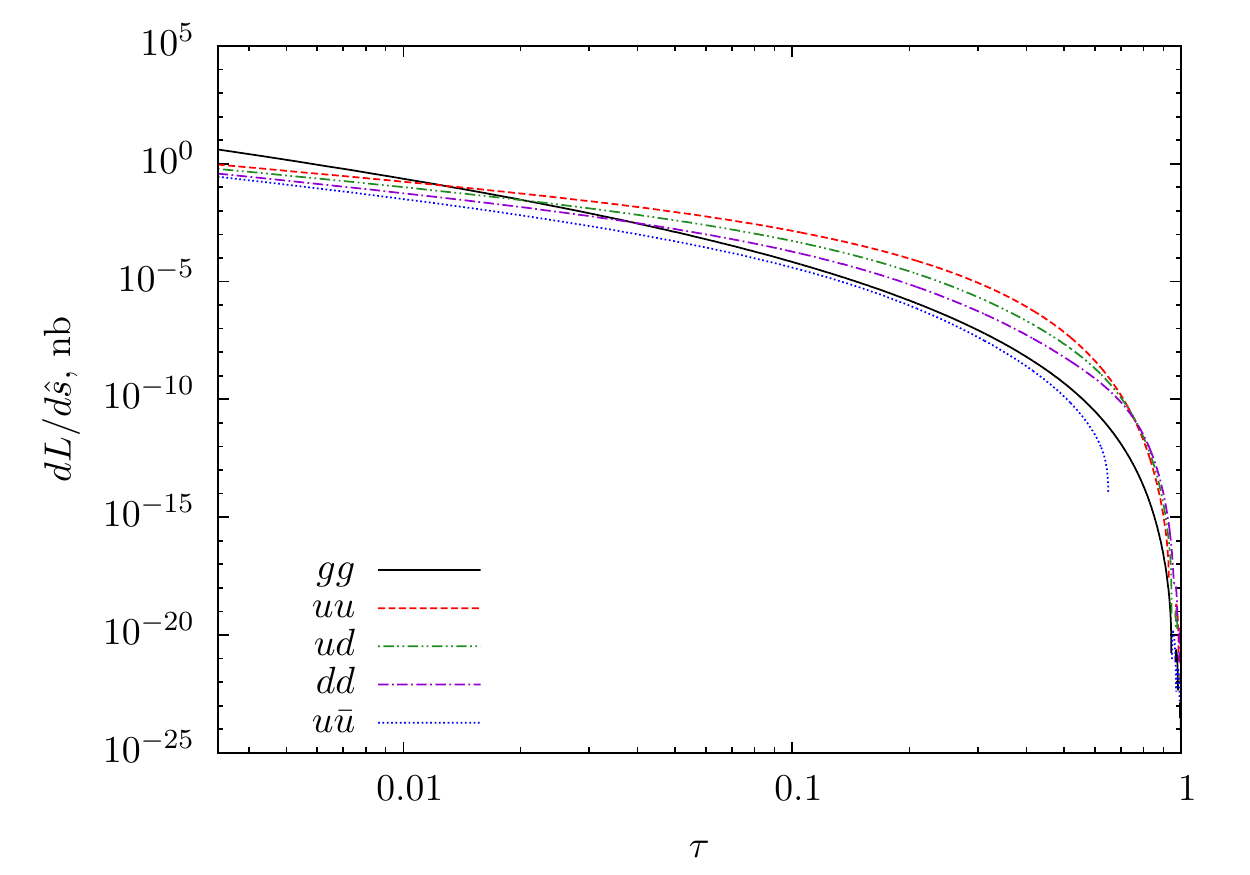}}
 \\
 \subfloat[Luminosities for $\sqrt{s} =  8$~TeV.]{\includegraphics{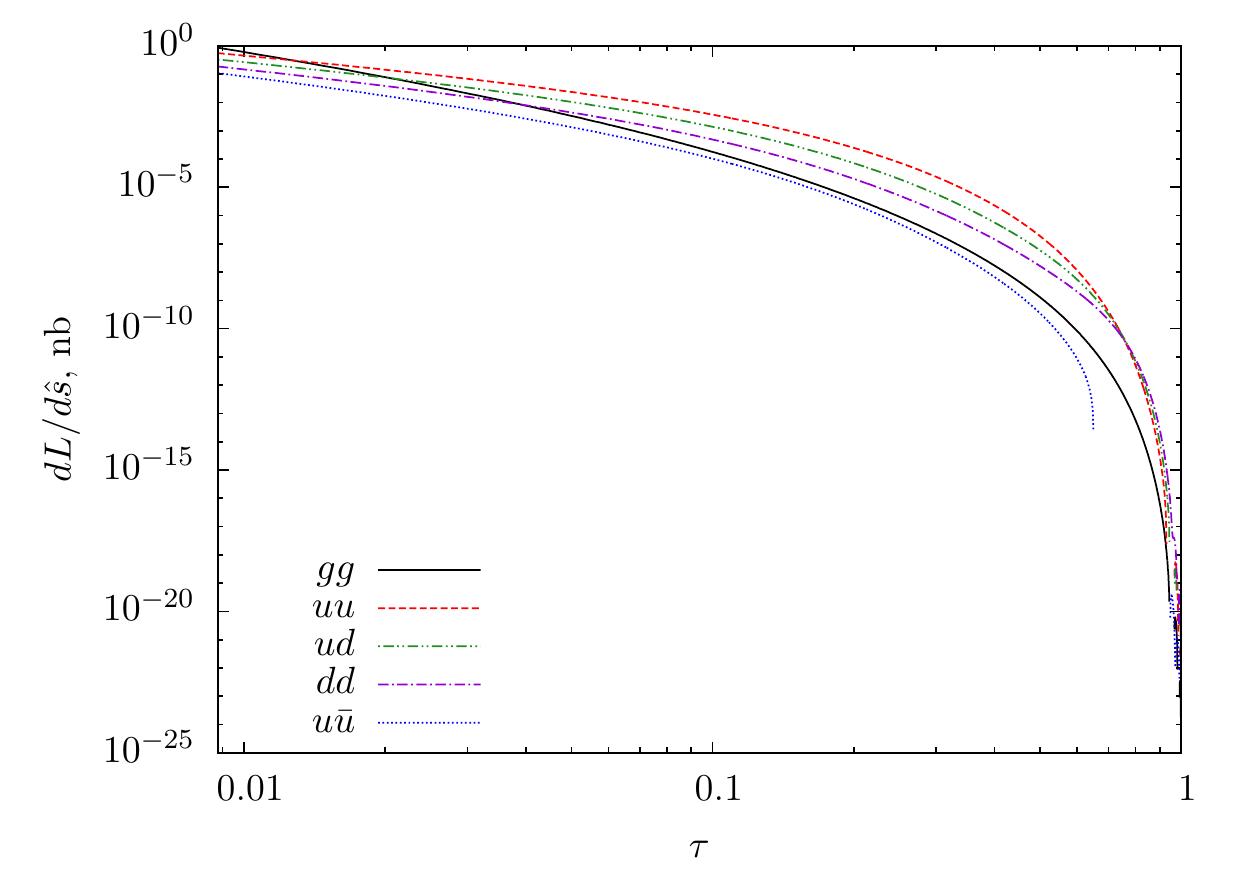}}
 \caption{Luminosities~\eqref{gg-luminosity}, \eqref{qq-luminosity} for
 gluon-gluon, $uu$, $ud$, $dd$ and $u \bar u$ collisions at $Q^2 = (750\text{
 GeV})^2$.}
 \label{fig:luminosity}
\end{figure}

Let us suppose that heavy leptons $L_i$ which couple to $S$ have electric
charges $Q_L$, and there are $N$ such degenerate leptons. The lagrangian is
similar to that of the heavy quarks case~\eqref{lagrangian-t}:
\begin{equation}
 \Delta \mathcal{L}
 = \tfrac12 (\partial_\mu S)^2
 - \tfrac12 m_S^2 S^2
 + \bar L_i \gamma_\mu (\partial_\mu - i g' \tfrac{Y_L}{2} B_\mu) L_i
 + m_L \bar L_i L_i
 + \lambda_L \bar L_i L_i S,
\end{equation}
where we assume equal lepton masses and $S$ couplings.  For $S \to \gamma
\gamma$ width we obtain:
\begin{equation}
 \Gamma_{S \to \gamma \gamma}
 = \left( \frac{\alpha}{3 \pi} \right)^2
   (N Q_L^2)^2
   \frac{m_S^3 \lambda_L^2}{16 \pi m_L^2}
   \lvert F(\beta) \rvert^2, \ 
   \beta = \left( \frac{2 m_L}{m_S} \right)^2.
\end{equation}
Production of $S$ at the LHC occurs through fusion of two virtual photons
emitted by quarks which reside in the colliding protons. Let us estimate
the production cross section. For the partonic cross section we get:
\begin{equation}
 \sigma_{q_1 q_2 \to q_1 q_2 S}^{(\gamma \gamma)}(\hat s)
 = \frac{8 \alpha^2}{m_S^3} e_1^2 e_2^2 \, \Gamma_{S \to \gamma \gamma}
   \left[
        f \left( \frac{m_S^2}{\hat s} \right)
        \left(
         \ln \left( \frac{m_L^2 \hat s}{\Lambda_\text{QCD}^2 m_S^2}
        \right)
      - 1 \right)^2
    - \frac13 \ln^3 \left( \frac{\hat s}{m_S^2} \right)
   \right],
 \label{qq->qqS-xsection}
\end{equation}
where $e_1$ and $e_2$ are charges of the colliding quarks, $\hat s = x_1 x_2 s
\equiv \tau s$ is the invariant mass of the colliding quarks, and $f(z)$ is
given by~\eqref{ee->eeS-f}. We should multiply~\eqref{qq->qqS-xsection} by
quark distribution functions and integrate over $x_1$ and $x_2$:
\begin{equation}
 \sigma_{pp \to SX}^{(\gamma \gamma)}(s)
 = \sum\limits_{q_1, q_2}
    \;
    \int\limits_{m_S^2 / s}^1
      \sigma_{q_1 q_2 \to q_1 q_2 S}^{(\gamma \gamma)}(\tau s)
    d \tau
    \cdot s \cdot \frac{dL_{q_1 q_2}}{d \hat s}(Q^2, \tau),
\end{equation}
where the sum should be performed over valence $uu$, $ud$, $du$, and $dd$ quark collisions,
and sea quarks should be taken into account as well.\footnote{
  $uu$ contribution constitutes 50\% of the cross section at $\sqrt{s} = 13$~TeV
  with another 24\% coming from $ud$ and $\bar u u$.
} Quark luminosity equals:
\begin{equation}
 \frac{d L_{q_1 q_2}}{d \hat s}(Q^2, \tau)
 = \frac{1}{s}
   \int\limits_{\ln \sqrt{\tau}}^{-\ln \sqrt{\tau}}
    q_1(x_1, Q^2) q_2(x_2, Q^2)
   d y,
 \label{qq-luminosity}
\end{equation}
$x_1 = \sqrt{\tau} \e^y$, $x_2 = \sqrt{\tau} \e^{-y}$. We take $Q^2 = m_S^2$ and
use quark distributions from~\cite{mmht}. Quark and gluon luminosity functions
for $s = 13$~TeV and $s = 8$~TeV are shown in Fig.~\ref{fig:luminosity}.

Cross sections in the case of one heavy lepton with charge $Q_L = 1$, Yukawa
coupling constant $\lambda_L = 2$ and mass $m_L = 400$~GeV are shown in
Table~\ref{lepton-table}. For $\Lambda_\text{QCD} = 300$~MeV and $\sqrt{s} =
13$~TeV we get $\sigma_{pp \to SX}^{(\gamma \gamma)} \approx 11$~ab,\footnote{
 According to Eq.~(12) from the recent paper~\cite{ryskin}, this cross section
 equals 25~ab.
} while the experimental result~\eqref{pp->SX-experimental-value} is three
orders of magnitude larger. We come to the conclusion that $\sum N Q_L^2 \approx
30$ is needed: we need either 30 leptons with unit charges, or one lepton with
charge 6, or several multicharged leptons.\footnote{
 If $\sigma_{pp \to SX}^{(\gamma \gamma)} = 25$~ab, then 30 should be replaced
 with 20.
}

It is natural to suppose that leptons with charge one mix with the Standard
Model leptons and become unstable. Search for such particles was performed at
the LHC, and the lower bound $m_L > 170$~GeV was
obtained~\cite{atlas-heavy-lepton}. See also~\cite{ellis}, where bounds on
masses and mixings of $L$ are discussed. For masses above 200~GeV the existence
of $L$ is still relatively unconstrained.

Cross section for quasielastic $S$ production can be estimated with the help of
the following equation:
\begin{equation}
 \sigma_{pp \to ppS}
 = \frac{8 \alpha^2}{m_S^3} \Gamma_{S \to \gamma \gamma}
   \left[
      f \left( \frac{m_S^2}{s} \right)
      \left( \ln \left( \frac{s}{m_S^2} \right) - 1 \right)^2
    - \frac13 \ln^3 \left( \frac{s}{m_S^2} \right)
   \right].
\end{equation}
For $\sqrt{s} = 13$~TeV, $\lambda_L = 2$ and $m_L = 400$~GeV it equals
4.1~ab.\footnote{
 According to Eq.~(24) from~\cite{ryskin}, quasielastic cross section is two times
 smaller.
}

\renewcommand{\arraystretch}{1.5}
\begin{table}[h]
 \caption{Cross sections (in ab) for double photon production in the leptophilic
  model for different values of $\Lambda_\text{QCD}$ and proton collision
  energies.}
 \label{lepton-table}
 \centering
 \begin{tabular}{|cc|ccc|}
  \cline{3-5}
  \multicolumn{1}{c}{}
  && \multicolumn{3}{c|}{$\Lambda_\text{QCD}$, GeV} \\
  \multicolumn{1}{c}{}
  &   & $ 0.1$ & $ 0.3$ & $ 1.0$ \\ \hline
  \multirow{3}{*}{\rotatebox{90}{$\sqrt{s}$, TeV}}
  & 7 & $ 2.5$ & $ 1.9$ & $ 1.3$ \\
  & 8 & $ 3.8$ & $ 2.9$ & $ 2.0$ \\
  &13 & $ 15 $ & $ 11 $ & $ 7.8$ \\
  \hline
 \end{tabular}
\end{table}
\renewcommand{\arraystretch}{1.0}

\section{Conclusions}

We analyze the possibility that the enhancement at 750~GeV diphoton invariant mass
observed by ATLAS and CMS is due to decays of a new scalar $S$. We found that
production of $S$ in gluon fusion in a minimal model with one additional heavy
Dirac quark $T$ can have value of $\sigma_{pp \to SX} \cdot \Br(S \to \gamma
\gamma)$ compatible with data. An upper bound on the mixing of $S$ with $h(125)$
is obtained. If heavy leptons $L$ are introduced instead of
$T$, then $S$ can be produced at LHC in photon fusion.  However, in order to
reproduce experimental data many leptons $L_i$ are needed and\slash{}or they
should be multicharged. If the existence of $S$ will be confirmed by future data
then production of heavy vector-like quarks and\slash{}or leptons at the LHC
should be looked for. The search for $S \to Z \gamma, ZZ, WW$ and $hh$ would be
also of great importance.

S.~G., M.~V. and E.~Zh. are partially supported under the grants RFBR No.
14-02-00995 and 16-02-00342, and by the Russian Federation Government under the
grant NSh-6792.2016.2. S.~G. and E.~Zh. are also supported by MK-4234.2015.2 and
16-32-00241. In addition, S.~G. is supported by RFBR under grants 16-32-60115,
by Dynasty Foundation and by the Russian Federation Government under Grant No.
11.G34.31.0047.


\begin{thebibliography}{9}
 \bibitem{atlas-750gg}
  The ATLAS collaboration,
  ATLAS-CONF-2015-081 (2015).
 \bibitem{cms-750gg}
  The CMS collaboration,
  CMS-PAS-EXO-15-004 (2015).
 \bibitem{mmht}
  L.~A.~Harland-Lang, A.~D.~Martin, P.~Motylinkski, R.~S.~Thorne,
  Eur. Phys. J. {\bf C}75 (2015) 204; arXiv:1412.3989.
 \bibitem{djouadi}
  J.~Baglio, A.~Djouadi,
  JHEP 1103 (2011) 055; arXiv:1012.0530.
 \bibitem{harlander}
  R.~V.~Harlander, W.~Kilgol.
  Phys. Rev. Lett. 88 (2002) 201801.
 \bibitem{buckley}
  M.~R.~Buckley,
  arXiv:1601.04751.
 \bibitem{isosinglet}
  S.~I.~Godunov, A.~N.~Rozanov, M.~I.~Vysotsky, E.~V.~Zhemchugov,
  Eur. Phys. J. C 76 (2016) 1;
  arXiv:1503.01618.
 \bibitem{atlas-4l}
  The ATLAS collaboration,
  ATLAS-CONF-2015-59.
 \bibitem{franceschini}
  R.~Franceschini {\it et. al.},
  arXiv:1512.04933.
 \bibitem{atlas-zz}
  The ATLAS collaboration,
  Eur. Phys. J. C76 (2016) 1, 45;
  arXiv:1507.05930.
 \bibitem{atlas-hh}
  The ATLAS collaboration,
  Phys. Rev. D92 (2015) 092004;
  arXiv:1509.04670.
 \bibitem{cms-8TeV-bound}
  The CMS collaboration,
  Phys. Lett. {\bf B}750 (2015) 494; arXiv:1506.02301.
 \bibitem{tt-bound-1}
  The ATLAS collaboration,
  JHEP 1510 (2015) 150; arXiv:1504.04605.
 \bibitem{tt-bound-2}
  The ATLAS collaboration,
  JHEP 1508 (2015) 105; arXiv:1505.04306.
 \bibitem{tt-bound-3}
  The CMS collaboration,
  Phys. Rev. {\bf D}93 (2016) 012003; arXiv:1509.04177.
 \bibitem{buchkremer}
  M.~Buchkremer,
  Proc. of 49th Rencontres de Moriond on Electroweak Interactions and Unified
  Theories (2014), p.~519; arXiv:1405.2586.
 \bibitem{dijet-background}
  The ATLAS collaboration,
  Phys. Rev. {\bf D}91 (2015) 052007; arXiv:1407.1376.
 \bibitem{pdg}
  K.~A.~Olive {\it et al.} (Particle Data Group).
  Chin. Phys. C, 2014, {\bf 38}(9): 090001.
 \bibitem{ginzburg}
  V.~M.~Budnev, I.~F.~Ginzburg, G.~V.~Meledin, V.~G.~Serbo,
  Phys. Reports 15 (1975) 181.
 \bibitem{ryskin}
  L.~A.~Harland-Lang, V.~A.~Khoze, M.~G.~Ryskin,
  arXiv:1601.07187.
 \bibitem{atlas-heavy-lepton}
  The ATLAS collaboration,
  JHEP 09 (2015) 108,
  arXiv:1506.01291.
 \bibitem{ellis}
  A.~Djouadi, J.~Ellis, R.~Godbole, J.~Quevillon,
  arXiv:1601.03696.
\end{thebibliography}
\end{document}